\documentclass[aps,prb,reprint, amsmath, amssymb]{revtex4-1}

\usepackage{bm}
\usepackage{booktabs}
\usepackage[retainorgcmds]{IEEEtrantools}
\usepackage{graphicx}
\usepackage{mathrsfs}
\usepackage{amsmath}
\usepackage{amssymb}
\usepackage{color}
\usepackage{amsfonts}

\newcommand{\ud}{\,\mathrm{d}}

\DeclareMathOperator{\tr}{tr}

\begin{document}

\title{Quantum thermodynamics and work fluctuations with applications to magnetic resonance}
\date{\today}
\author{Wellington L. Ribeiro}
\author{Gabriel T. Landi}
\email{gtlandi@gmail.com}
\author{Fernando L. Semi\~ao}
\affiliation{Universidade Federal do ABC,  09210-580 Santo Andr\'e, Brazil}

\begin{abstract}

In this paper we give a pedagogical introduction to the ideas of quantum thermodynamics and work fluctuations, using only basic  concepts from  quantum  and statistical mechanics. 
After reviewing the concept of work, as usually taught in thermodynamics and statistical mechanics, we discuss the framework of non-equilibrium processes in quantum systems together with some modern developments, such as the Jarzynski equality and its connection to the second law of thermodynamics. 
We then apply these results to the problem of magnetic resonance, where all calculations may be done exactly. 
It is shown in detail how to build the statistics of the work, both for a single particle and for a collection of non-interacting particles. 
We hope that this paper may serve as a tool to bring the new student up to date on the recent developments in non-equilibrium  thermodynamics of quantum systems.

\end{abstract}
\maketitle{}

%
%
%
%
\section{\label{sec:int}Introduction}

Thermodynamics was initially developed to deal with macroscopic systems\cite{Fermi1956,Callen1985} and is thus based on the idea that a handful of macroscopic variables, such as volume, pressure and temperature, suffice to completely characterize a system. 
However, after the advent of the atomic theory, it became clear that the variables of the underlying microscopic world are constantly fluctuating  due to the inherent chaos and randomness of the micro-world. 
Statistical mechanics was thus developed as a theory connecting these microscopic fluctuations with the emergent macroscopic variables. 
Since one usually deals with a large number of particles, the relative fluctuations become negligible, so that thermodynamic measurements usually coincide very well with expectation values of the microscopic fluctuating quantities (which is a consequence of the  law of large numbers\cite{Ross2010}). 

Equilibrium statistical mechanics is now a very well established and successful theory. 
Its main result is the Gibbs formula for the canonical ensemble\cite{Gibbs,Feyman1998} which provides a fundamental bridge between microscopic physics 
and thermodynamics for any equilibrium situation. 
Conversely, far less is known about non-equilibrium processes. 
The reason is that in this case the handful of parameters used in thermodynamics no longer suffice, forcing one to know  the full dynamics of the system;
i.e., one must study Newton's or Schr\"odinger's equation for all constituent particles,  thus making the problem much more difficult.

These difficulties led researchers to look for non-equilibrium processes in the realm of small systems. 
On the one hand, in these systems the dynamics are somewhat easier to describe since there are fewer particles.
But on the other hand, fluctuations become important and must therefore be included in the description. 
Substantial progress in experimental methods also made possible for the first time to experiment with small systems, and therefore also contributed to this shift.

The random fluctuations present in small systems also affect thermodynamic quantities such as work and heat. 
A beautiful example is the use of optical tweezers to fold and unfold individual RNA molecules.\cite{Liphardt2002}
The molecules are immersed in water and therefore subject to the incessant  fluctuations of brownian motion. 
Thus, the amount of work required to fold a molecule will be different each time we repeat the experiment and should therefore be interpreted as a random variable.
The same is true of the work that is extracted from the molecule when it is unfolded. 
In some realizations, it is even possible to extract work without any changes in the thermodynamic state of the system, something which would contradict the second law of thermodynamics. 

This introduces the idea that fluctuations in small systems could lead to \emph{local violations} of the second law. 
These  violations were first observed in  fluid simulations in the beginning of the 1990s by Evans, Cohen, Gallavotti and  collaborators. \cite{Evans1993,Gallavotti1995b} 
Afterwards, in 1997 and 1998 came two very important (and intimately related) breakthroughs by  Jarzynski\cite{Jarzynski1997,Jarzynski1997a} and  Crooks.\cite{Crooks1998,Crooks2000}
They showed that the work performed in a non-equilibrium (e.g. fast) process, when interpreted as a random variable, obeyed a set of exact relations that touched deeply into the nature of irreversibility and the second law. 
Nowadays these relations go by the generic name of  \emph{fluctuation theorems} (not to be confused with the fluctuation-dissipation theorem) and can be interpreted as generalizations of the second law for fluctuating systems.
These theoretical findings were later confirmed by  experiments in diverse areas, such as trapped brownian particles,\cite{Martinez2015, Gomez-Solano2011}
mechanical oscillators\cite{Douarche2005} and biological systems. \cite{Collin2005, Liphardt2002}

With time, researchers began to look for similar results in quantum systems, both for unitary\cite{Mukamel2003,Talkner2007} as well as for open\cite{Crooks2008, Talkner2009} quantum dynamics.
Here, in addition to the thermal fluctuations, one also has intrinsically quantum fluctuations, leading to a much richer platform to work with.
One may therefore ask to what extent will these quantum fluctuations affect thermodynamic quantities such as heat and work.
 This research was also motivated by the remarkable progress of the past decade in the experimental control of atoms and photons, particularly in  areas such as magnetic resonance, ultra-cold atoms and quantum optics.
In fact, the first experimental verifications of the fluctuation theorems for quantum systems were done only recently, using nuclear magnetic resonance\cite{Batalhao2014} and trapped ions. \cite{An2014}

It is a remarkable achievement of modern day physics that we are now able to test the thermodynamic properties of systems containing only a handful of particles. 
But despite being an active area of research, the basic concepts in this field can be understood using only undergraduate quantum and statistical mechanics. 
The purpose of this paper is to provide an introduction to some of these concepts in the realm of quantum systems. 
Our main goal is to introduce the reader to the idea that work may be treated as a random variable.
We then show how to construct all of its  statistical properties, such as the corresponding probability distribution of work or the characteristic function. 
We discuss how these quantities may be used to derive Jarzynski's equality, \cite{Jarzynski1997,Jarzynski1997a} thence serving as a quantum mechanical derivation of the second law.
To illustrate these new ideas, we  apply our results to the problem of magnetic resonance, an elegant textbook example that can be worked out analytically.

%
%
%
%
\section{\label{sec:thermo}Work in thermodynamics and statistical mechanics}

\subsection{Thermodynamic description}

Consider any physical system described by a certain Hamiltonian $H$.
When this system is placed in contact with a heat reservoir, energy may flow between the system and the bath. 
This change in energy is  called \emph{heat}. 
But the energy of a system may also change by means of an external agent, which manually changes some parameter in the Hamiltonian of the system. 
These types of changes are called \emph{work}. 

Heat and work are not properties of the system. 
Rather, they are the outcomes of processes which alter the state of the system. 
If during a certain interval of time an amount of heat $Q$  entered the system and a work $W$ was performed on the system, energy conservation implies that the total energy  $U$ of the system must have changed by 
\begin{equation}\label{first_law}
\Delta U = Q + W,
\end{equation}
which is the first law of thermodynamics. When $W>0$ we say the external agent performed work on the system. 
Conversely, when $W<0$ we say the system performed work on the external agent (for instance, when a gas expands a piston).

The process of performing work on a system may be described microscopically in a very general way through  the change of  some parameter $\lambda$ in the Hamiltonian.
We shall call this the \emph{work parameter}.
Examples include the volume of a container, an electric or magnetic  field, or the stiffness of a harmonic trap. 
In order to describe exactly how the work was performed we must also specify the  \emph{protocol} to be used. 
This means specifying exactly under what conditions  are the changes being made and with what time-dependence $\lambda(t)$  is changing. 
We usually assume that the process lasts between a time $t=0$ and a time $t = \tau$, during which $\lambda$ varies in some pre-defined way   from an initial value $\lambda_i = \lambda(0)$ to a final value $\lambda_f = \lambda(\tau)$.

Overall, describing an arbitrarily fast process can be a very difficult task since it requires detailed information about the dynamics of the system and how it is coupled to the bath. 
Instead, thermodynamics usually focuses on \emph{quasi-static} processes, in which $\lambda$ changes very slowly in order to ensure that throughout the process the system is always in thermal equilibrium.
Quasi-static processes have the advantage of being \emph{atemporal}: we do not need to specify the function $\lambda(t)$, but merely its initial and final value.
In this paper our focus will be on non-equilibrium processes. 
But before we can get there, we must first have a solid understanding of quasi-static processes.

Perhaps the most important example of a quasi-static process is the \emph{isothermal process}, in which the temperature of the system is kept constant throughout the protocol. 
Since work usually changes the temperature of an isolated system, to ensure a constant temperature  the system must remain coupled to a heat reservoir kept at a constant temperature $T$. 
It is also important to note that an isothermal process must \emph{necessarily} be quasi-static. 
For, if the process is not quasi-static, the temperature will not remain constant nor homogeneous. 
In fact, intensive quantities such as temperature and pressure are only defined in thermal equilibrium, so any process where these quantities are kept fixed must be quasi-static. 
Thus, henceforth whenever we speak of an isothermal process, it will already be implied that it is quasi-static.

According to Eq.~(\ref{first_law}), in an isothermal process not all work will be converted into useful energy for the system, since part is consumed as heat. 
The remainder is called the \emph{free energy}:
\begin{equation}\label{WF}
W = \Delta F = \Delta U - Q.
\end{equation}
where $\Delta F = F(T,\lambda_f) - F(T,\lambda_i)$ (cf. chapter 2 of Ref.~\onlinecite{Schrodinger} or  Ref.~\onlinecite{Fermi1956}). 
The energy is ``free''   because it is the part of $\Delta U$ that may be used to perform work.

\begin{figure}[!h]
\centering
\includegraphics[width=0.45\textwidth]{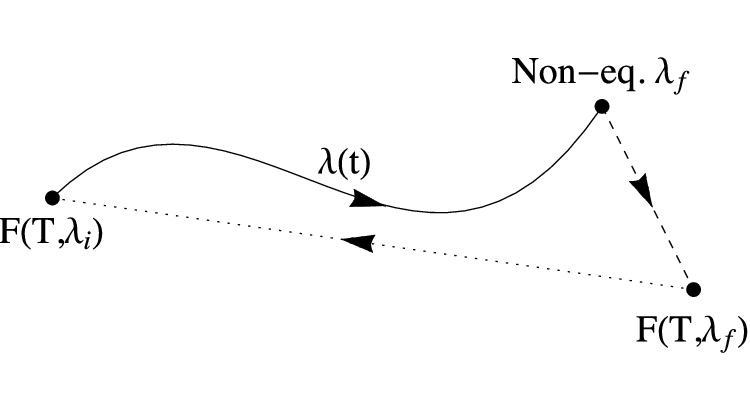}
\caption{\label{fig:diagram}
Diagram of a non-equilibrium process. 
Through the protocol $\lambda(t)$, the system is taken from an initial state $F(T,\lambda_i)$ to a final non-equilibrium state with parameter $\lambda_f$ (solid line). 
After the process is done, the system will eventually relax from the non-equilibrium state to the equilibrium state $F(T,\lambda_f)$ (dashed line).
Finally, the dotted line represents the journey back to the original state. 
}
\end{figure}

Now suppose we try to repeat the  process  above, but we do so too quickly, so that it cannot be considered quasi-static. 
The initial state is still $F(T,\lambda_i)$, but the final state will not be $F(T,\lambda_f)$.
Instead, the final state will be something complicated which depends exactly on how the process was performed (see Fig.~\ref{fig:diagram}). 
Notwithstanding, since the system is coupled to a bath,  if we leave it alone after the protocol is over,  it will eventually relax to the state $F(T,\lambda_f)$. 
Thus, overall, a certain amount of work $W$ was performed to take the system from $F(T,\lambda_i)$ to $F(T,\lambda_f)$. 
But this work is not equal to $\Delta F$, since Eq.~(\ref{WF}) holds only for quasi-static processes. 

Instead, according to the \emph{second law of thermodynamics}, the work done in the non-equilibrium process must always be larger or equal to $\Delta F$:
\begin{equation}\label{second_law}
W \geq \Delta F,
\end{equation} 
with the equality holding only for the isothermal (i.e. quasi-static) process.
This is a very important result. 
It means that if we wish to change the free energy of a system by $\Delta F$, the \emph{minimum} amount of work we need to perform is $\Delta F$ and will be accomplished in an isothermal (quasi-static) process. 
Any other protocol will require more work. 
The difference $W_\text{irr} = W - \Delta F \geq 0$, known as the \emph{irreversible work}, therefore represents the extra work that had to be done due to the particular choice of protocol.

The  inequality~(\ref{second_law}) can be interpreted as a direct consequence of the Kelvin statement of the second law:  \emph{``A transformation whose only final result is to transform into work, heat extracted from a source which is at the same temperature throughout, is impossible''}.\cite{Fermi1956}
The key part of this statement is the expression  ``only final result''. 
It means that it is impossible  to extract work from a  bath at a fixed temperature, without changing anything else (such as e.g. the thermodynamic state of the system). 
Extracting work while changing the thermodynamic state of a system is not a problem. 
For instance, we can use a heat source to make a gas expand and thence extract work from the expansion. 
But by the end of the process we will have altered the state of the gas, so the extraction of work was not the only outcome. 
What the second law says is that it is impossible to extract work from a single source at a fixed temperature \emph{and} keep the state of the system intact. 

To make the connection with Eq.~(\ref{second_law}), consider a process divided in 3 steps, represented by the three lines in Fig.~\ref{fig:diagram}. 
In the first step we perform a certain amount of work $W$ in a non-equilibrium process. 
In the second, we perform no work and allow the system to relax from the non-equilibrium state to $F(T,\lambda_f)$. 
Finally, in the third process (represented by a dotted line in Fig.~\ref{fig:diagram}), we go back \emph{quasi-statically} from $F(T,\lambda_f)$ to $F(T,\lambda_i)$.
The amount of work required for the return journey is $W_\text{return} = - \Delta F$ because we assume that this part was quasi-static.
In the end, we are back to the original state, having performed a total work $W + W_\text{return} = W - \Delta F$. 
According to the second law, this total work cannot be negative, because that would mean we would have extracted work from a reservoir at a fixed temperature, without any changes in the state of the system. 
Consequently, $W - \Delta F \geq 0$, which is  Eq.~(\ref{second_law}).

\subsection{\label{ssec:infinitesimal}Isothermal processes in equilibrium statistical mechanics}

Let us now analyze the isothermal process quantitatively. 
Since it is a quasi-static process, we may always decompose it into a series of \emph{infinitesimal processes}, where $\lambda$ is changed slightly to $\lambda + \ud \lambda$. The full process is then simply a succession of these small steps.

We assume that initially we had a system with Hamiltonian $H(\lambda) = H$ in thermal equilibrium with a heat bath at a temperature $T$. 
According to statistical mechanics, its state is then given by the Gibbs density operator
\begin{equation}\label{rho_thermal}
\rho_\text{th} = \frac{e^{-\beta H}}{Z},
\end{equation}
where $Z = \tr(e^{-\beta H})$ is the partition function and $\beta = 1/T$ (we choose Boltzmann's constant $k_B$ equal to unity). 
We may also write this expression in terms of the energy eigenvalues $E_n$ and eigenvectors $|n\rangle$. 
The probability of finding the system in the state $|n\rangle$ is therefore, 
\begin{equation}\label{prob}
P_n = \langle n | \rho_\text{th} | n \rangle = \frac{e^{-\beta E_n}}{Z}.
\end{equation}
Moreover, the internal (average) energy of the system may be written as
\begin{equation}\label{U}
U = \langle H \rangle = \tr(H \rho_\text{th}) = \sum\limits_n E_n P_n
\end{equation}

When we change $\lambda$ to $\lambda + \ud \lambda$, both $E_n$ and $P_n$ will change.
Hence, $U$ will change by:
\begin{equation}\label{du}
\ud U = \sum\limits_n \ud (E_n P_n) = \sum\limits_n \big[(\ud E_n) P_n + E_n (\ud P_n) \big].
\end{equation}
This separation of $\ud U$ in two terms allows for an interesting physical interpretation.\cite{Crooks1998} 

The change in $\lambda$ is infinitesimal and instantaneous.
So immediately after this change, the system has not yet responded to it. 
This corresponds to the first term: it is the average of the energy change  $\ud E_n$ over the old (unperturbed) probabilities $P_n$. 
In the second term, the energy is fixed and the probabilities change. 
We interpret this as the second step, where the system adjusts itself with the bath in order to return to equilibrium. 
Thus, each infinitesimal process may be separated in two parts. 
The first part is the work performed and the second is the heat exchanged as the system relaxes to equilibrium. 

This motivates us to define:
\begin{IEEEeqnarray}{rCl}
\label{dW} \delta W &=& \sum\limits_n (\ud E_n) P_n, \\[0.2cm]
\label{dQ} \delta Q &=& \sum\limits_n E_n (\ud P_n),
\end{IEEEeqnarray}
so that Eq.~(\ref{du}) may be written as $\ud U = \delta Q + \delta W$ (we use $\delta$ instead of $\ud$ simply to emphasize that heat and work are not exact 
differentials.\cite{Callen1985})
In order to  better understand the physical meaning of these formulas, we will now explore them in more detail.

We start with $\delta W$ and  show that it is related to the free energy of the system, defined as
\begin{equation}\label{F}
F = - T \ln Z.
\end{equation}
To see this, first note that since the temperature $T$ is fixed, $\ud F = - T \ud Z/Z$. 
Since $Z = \sum_n e^{-E_n/T}$, one may then readily show that $\ud F = \sum_n (\ud E_n) P_n$, which is precisely Eq.~(\ref{dW}). Thus
\begin{equation}
\delta W = \ud F.
\end{equation}
From this result, Eq.~(\ref{WF}) is recovered by integrating over the several infinitesimal steps. 

Next we turn to  $\delta Q$ in Eq.~(\ref{dQ}). 
Instead of trying to manipulate $\ud P_n$, we can use  the following neat trick: invert Eq.~(\ref{prob}) to write  $E_n = - T \ln (Z P_n)$. 
If we substitute this in Eq.~(\ref{dQ}) we get two terms, one proportional to $\ln(Z)$ and the other to $\ln(P_n)$. The term with $\ln(Z)$ will be 
\begin{equation}
-T \sum\limits_n \ln(Z) \ud (P_n) = -T \ln(Z) \ud \left(\sum\limits_n P_n\right) = 0,
\end{equation}
since $\sum_n P_n = 1$ and $\ud (1) = 0$. 
Thus, we are left only with 
\begin{equation}
\delta Q = -T \sum\limits_n (\ud P_n) \ln P_n.
\end{equation}
But now note that, by the chain rule,
\begin{equation}
\ud \left( \sum\limits_n P_n \ln P_n \right) = \sum\limits_n (\ud P_n) \ln P_n + \sum\limits_n \frac{P_n}{P_n} (\ud P_n),
\end{equation}
and the last term is also zero for the same reason as above. 
Hence, we conclude that 
\begin{equation}
\delta Q = - T \ud \left( \sum\limits_n P_n \ln P_n \right).
\end{equation}
We see that even though $\delta Q$ is not a function of state, it is related to the variation of a quantity which is a function of state. 
We define the \emph{entropy} as 
\begin{equation}\label{entropy}
S = - \sum\limits_n P_n \ln P_n,
\end{equation}
so that we finally arrive at 
\begin{equation}\label{QS}
\delta Q = T \ud S
\end{equation}
This relation, we emphasize, holds only for infinitesimal processes. 
For finite and irreversible processes, there may be additional contributions to the change in entropy. 

We therefore see that it is possible to give microscopic definitions to thermodynamic quantities such as heat and work. Moreover, 
it is possible to relate them to functions of state which can be constructed from the initial density matrix $\rho_\text{th}$.
While these thermodynamic quantities may be defined independently of statistical mechanics, we believe that this microscopic description helps to clarify their physical meanings.

%
%
%
%
\section{\label{sec:work}Work as a random variable}

In 1997 Jarzynksi discovered that great insight into the properties of non-equilibrium processes could be gained by treating 
work as a \emph{random variable}.\cite{Jarzynski1997,Jarzynski1997a}
The idea is as follows. 
Consider the process where a movable piston is used to compress   a gas contained in a cylinder.
Since the molecules of the gas are moving chaotically and hitting the walls of the piston in all sorts of different ways, 
each time we press the piston, the gas molecules will exert back on us a different force. 
This means that the work needed to achieve a given compression will change each time we repeat the experiment. 
Hence, \emph{work may be treated as a random variable}. 

Of course, one may object that in most systems these fluctuations are negligibly small. 
But that does not stop us from interpreting $W$ in this way.
And, as we will soon see, this does bring several advantages. 
On the other hand, when dealing with microscopic systems, this interpretation becomes essential since fluctuations become significant.
A famous example is the work performed when folding RNA molecules,\cite{Liphardt2002,Collin2005} already discussed in Sec.~\ref{sec:int}.  

In addition to thermal fluctuations, some microscopic systems also have a strong contribution from quantum fluctuations. 
These fluctuations are related to the fact that in order to access the amount of work performed in a system, one must measure its energy and therefore collapse the wave-function. 
As is known from quantum mechanics, when this is done the system may tend to different states with different probabilities [cf. Eq.~(\ref{prob})].  
Thus, in quantum systems both thermal and quantum fluctuations must be taken into consideration.

\subsection{The Jarzynski equality}

Usually, our knowledge of non-equilibrium processes is restricted only to inequalities  such as  Eq.~(\ref{second_law}). 
The contribution of Jarzynski\cite{Jarzynski1997,Jarzynski1997a}  was to show that by interpreting $W$ as a random variable, one may obtain an \emph{equality}, even for a process performed arbitrarily far from equilibrium. 

Consider several realizations of a non-equilibrium process, such as that described by the solid line in Fig.~\ref{fig:diagram}. 
At each realization we always prepare the system in the same initial state. 
We then execute the protocol and measure the total work $W$ performed. 
After repeating this process many times, we may construct the probability distribution of the work, $P(W)$.
From this probability any average may be computed. For instance the average  work will be
\begin{equation}\label{Wave}
\langle W \rangle = \int P(W) \ud W
\end{equation}
Or we may study the average of other quantities such as $\langle W^2 \rangle$ and so on.

 Jarzynski's main result was to show that the statistical average of $e^{-\beta W}$  should satisfy:\cite{Jarzynski1997,Jarzynski1997a}
\begin{equation}\label{jarzynski}
\langle e^{-\beta W} \rangle  = e^{-\beta \Delta F}.
\end{equation}
where $\Delta F = F(T,\lambda_f) - F(T,\lambda_i)$, as described in Fig.~\ref{fig:diagram}.
This result is nothing short of remarkable. 
It holds for a process performed \emph{arbitrarily} far from equilibrium. 
And it is an \emph{equality}, which is a much stronger statement than the inequalities  we are used to in thermodynamics.
The appearance of $F(T, \lambda_f)$ in Eq.~(\ref{jarzynski}) is also quite surprising, since this is not the final state of the process.
It would only be the final state if the process were quasi-static. 
Instead, as indicated in Fig.~\ref{fig:diagram}, the final state is a non-equilibrium state which may differ substantially from $F(T,\lambda_f)$.
The appearance of $F(T,\lambda_f)$ therefore reflects the state that the system wants to go to, but cannot do so because the process is not sufficiently slow.

It is possible to show that the inequality~(\ref{second_law}) [ie, $W \geq \Delta F$] is contained within the Jarzynski equality. 
This can be accomplished using Jensen's inequality (see chapter 8 of Ref.~\onlinecite{Ross2010}), which states that $\langle e^{-\beta W} \rangle \geq e^{-\beta \langle W \rangle}$ 
Combining this with Eq.~(\ref{jarzynski}) then gives 
\begin{equation}\label{second_law2}
\langle W \rangle \geq \Delta F.
\end{equation}
We therefore see that, when we treat work as a random variable, the old results from thermodynamics are recovered for the average work. 

In macroscopic systems, by the law of large numbers (Ref.~\onlinecite{Ross2010}, chapter 8), individual measurements are usually very close to the average, so the distinction between the average work $\langle W \rangle$ and a single stochastic realization $W$ is  immaterial. 
But for microscopic systems this is usually not true. 
In fact, although  $\langle W \rangle \geq \Delta F$, the individual realizations $W$ may very well be smaller than $\Delta F$.
This therefore corresponds to \emph{local} violations of the second law. 
For large systems, these local violations become extremely rare. 
But for small systems, they may be measured experimentally.\cite{Liphardt2002,Collin2005,Batalhao2014}
If we know the distribution $P(W)$, then the probability of a local violation of the second law can easily be found as 
\begin{equation}\label{P_violation}
\text{Prob}(W < \Delta F) = \int\limits_{-\infty}^{\Delta F} P(W) \ud W
\end{equation}

The demonstration of the Jarzynski equality~(\ref{jarzynski}) requires detailed knowledge of the dynamics of the system; i.e., if it is classical, quantum, unitary, stochastic, etc.\cite{Note1}
 Below we will give a demonstration for one particular case of unitary quantum dynamics.

\subsection{Non-equilibrium unitary dynamics}

We will now consider in detail how to describe work in a non-equilibrium process. 
As mentioned above, this requires detailed knowledge of the dynamics of the system and how it is coupled to the heat bath. 
The last part is by far the most difficult since it requires concepts from open quantum systems, such as Lindblad dynamics or quantum Langevin equations.\cite{Breuer2007}
We will therefore make an important simplification and assume that the coupling to the heat bath is so weak that, during the protocol,  no heat is exchanged.
This situation is actually encountered very often in experiments since many systems are only weakly coupled to the bath. 
It  also simplifies considerably the description of the problem since it allows us to use Schr\"odinger's equation to describe the dynamics of the system.

We shall consider the protocol described in the previous section.
Initially the system had a Hamiltonian $H_i = H(\lambda_i)$ and was in thermal equilibrium with a bath at a temperature $T$. 
The initial state of the system is then given by the Gibbs thermal density matrix in Eq.~(\ref{rho_thermal}).
As a first step, we measure the energy of the system.
If we let $E_n^i$ and $|n\rangle$ denote the eigenvalues and eigenvectors of  $H_i$, then the energy $E_n^i$ will be obtained with probability $P_n = e^{-\beta E_n^i}/Z$ [cf. Eq.~(\ref{prob})].

Immediately after this measurement we initiate the protocol, changing $\lambda$ from $\lambda(0) = \lambda_i$ to $\lambda(\tau) = \lambda_f$ according to some pre-defined function $\lambda(t)$.  
If we assume that during this process the contact with the bath was very weak, we may assume that the state of the system will evolve according to:
\begin{equation}
|\psi(t)\rangle = U(t) |n\rangle
\end{equation}
where $U(t)$ is the unitary  time-evolution operator, which satisfies Schr\"odinger's equation (with $\hbar =1$): 
\begin{equation}\label{schrodinger}
i \partial_t U= H(t) U,\qquad U(0) = 1.
\end{equation}
For a derivation of this equation, see Ref.~\onlinecite{Sakurai2010}, chapter 2.

At the end of the process we measure the energy of the system once again. 
The Hamiltonian is now $H_f = H(\lambda_f)$ and therefore may have completely different energy levels $E_m^f$ and eigenvectors $|m\rangle$. 
The probability that we now measure an energy $E_m^f$ is 
\begin{equation}\label{conditional}
|\langle m | \psi(\tau) \rangle|^2 = |\langle m | U(\tau) | n\rangle|^2,
\end{equation}
which may be interpreted as the conditional probability that a system  initially in $|n\rangle$ will be found in $|m\rangle$ after a time $\tau$.

Since no heat may be exchanged with the environment, any change in the energy must necessarily be attributed to the work performed by the external agent. 
The energy obtained in the first measurement was $E_n^i$ and the energy obtained in the second measurement  was $E_m^f$. 
We  then define the work performed by the external agent as 
\begin{equation}\label{work_def}
W = E_m^f - E_n^i.
\end{equation}
Both $E_n^i$ and $E_m^f$ are fluctuating quantities which change during each realization of the experiment. 
The first energy $E_n^i$ is random due to thermal fluctuations and the second, $E_m^f$, is random due to quantum fluctuations. 
Consequently, $W$ will also be a random variable, encompassing both thermal and quantum fluctuations.

\subsection{Distribution of work and characteristic function}

We will now obtain an expression for the probability distribution $P(W)$ obtained by repeating the above protocol several times. 
This can be accomplished by noting that we are dealing here with a 2-step measurement processes. 
From probability theory, if $A$ and $B$ are two events, the total probability $P(A,B)$ that both events occur may be written as 
\begin{equation}
P(A,B) = P(A|B) P(B),
\end{equation}
where $P(B)$ is the probability that $B$ occurs and $P(A|B)$ is the conditional probability that $A$ occurs given that $B$ has occurred. 
In our context, $P(A|B)$ is given by Eq.~(\ref{conditional}) whereas $P(B)$ is simply the initial probabilities $P_n$.
Hence, the probability that both events have occurred is 
\begin{equation}
\text{Prob}(E_n^i \rightarrow \text{protocol} \rightarrow E_m^f)= |\langle m | U(\tau) | n\rangle|^2 P_n 
\end{equation}
Since we are interested in the work performed, we may then write 
\begin{equation}\label{PW}
P(W) = \sum\limits_{n,m} |\langle m | U(\tau) | n\rangle|^2 \; P_n \; \delta[W - (E_m^f - E_n^i)],
\end{equation}
where $\delta(x)$ is Dirac's delta function. 
This formula is perhaps best explained in words: we sum over all allowed events, weighted by their probabilities, and catalogue the terms according to the values of $E_m^f - E_n^i$.

Albeit exact, Eq.~(\ref{PW}) is not very convenient to work with. 
In most systems there are  a large number of allowed energy levels and therefore an even larger number of allowed energy differences $E_m^f - E_n^i$.
It is much more convenient to work with the \emph{characteristic function}, defined as the Fourier transform of the original distribution
\begin{equation}\label{G}
G(r) = \langle e^{i r W} \rangle = \int\limits_{-\infty}^\infty P(W) e^{i r W} \ud W.
\end{equation}
From $G(r)$ we may recover the original distribution from the inverse Fourier transform:
\begin{equation}\label{PG}
P(W) = \frac{1}{2\pi} \int\limits_{-\infty}^\infty \ud r G(r) e^{-i r W}
\end{equation}
Since $P(W)$ and $G(r)$ are Fourier transforms of each other, they both contain the same amount of information. 

With the help of Eq.~(\ref{PW}) we may write
\begin{IEEEeqnarray*}{rCl}
G(r)&=& \sum\limits_{n,m} |\langle m | U | n \rangle|^2 \; P_n e^{i r (E_m^f - E_n^i)},  \\[0.2cm]
	&=& \sum\limits_{n,m} \langle n| U^\dagger e^{i r E_m^f}| m \rangle \langle m | U e^{-i r E_n^i} P_n |n\rangle,  \\[0.2cm]
	&=& \sum\limits_{n,m} \langle n| U^\dagger e^{i r H_f}| m \rangle \langle m | U e^{-i r H_i} \rho_\text{th} |n\rangle\\[0.2cm]
	&=& \text{tr}\Big\{ U^\dagger(\tau) e^{i r H_f} U(\tau) e^{-i r H_i} \rho_\text{th} \Big\}.
\end{IEEEeqnarray*}
Hence, we conclude that 
\begin{equation}\label{G2}
G = \text{tr}\Big\{ U^\dagger(\tau) e^{i r H_f} U(\tau) e^{-i r H_i} \rho_\text{th} \Big\}.
\end{equation}
The characteristic function does not have  a particularly important physical meaning. 
But, since it is written as the trace of a product of operators, it is usually much more convenient to work with than $P(W)$. 
In many aspects, it plays a role somewhat similar to the partition function $Z$ in equilibrium statistical mechanics.
One usually does not question the physical meaning of $Z$, but rather uses it as a convenient quantity from which observables such as the energy and entropy may be extracted.

From $G(r)$ we may also readily extract the statistical moments of $W$.
To see this, we expand Eq.~(\ref{G}) in a Taylor series in $r$ to find
\begin{equation}\label{G_series}
G(r) = \langle e^{i r W}\rangle = 1 + i r \langle W \rangle - \frac{r^2}{2} \langle W^2 \rangle - i \frac{r^3}{3!} \langle W^3 \rangle + \ldots.
\end{equation}
Hence $\langle W^n \rangle$ will be proportional to the term of order $r^n$ in the expansion.

On the other hand, we may  obtain quantum mechanical formulas for the moments by doing a similar expansion in Eq.~(\ref{G2}). 
The average work, for instance, is found to be 
\begin{equation}\label{W11}
\langle W \rangle = \langle H_f\rangle_\tau - \langle H_i \rangle_0,
\end{equation}
where, given any operator $A$, we define
\begin{equation}\label{evolution_observable}
\langle A \rangle_t = \tr\bigg\{U^\dagger (t) A U(t) \rho_\text{th}  \bigg\}.
\end{equation}
as the expectation value of this operator at time $t$, a result which follows directly from the fact that the state of the system at time $t$ is $\rho(t) = U(t)\rho_\text{th} U^\dagger(t)$.
We therefore see that $\langle W \rangle$ is simply the difference between the average energy at time $\tau$ and the average energy at time $0$, which is intuitive.
One may continue with the expansion of eq.~(\ref{G2}) to obtain formulas for higher order moments. 
Unfortunately, they do not acquire such a simple form.

The characteristic function may also be sued to demonstrate the Jarzynski equality~(\ref{jarzynski}) since, based on the  definition~(\ref{G}),
we should have  $G(r=i\beta) = \langle e^{-\beta W} \rangle$. 
But from Eqs.~(\ref{rho_thermal}) and (\ref{G2}) we find that
\begin{equation}
G( i\beta) = \frac{1}{Z_i} \tr (U^\dagger e^{-\beta H_f} U) = \frac{1}{Z_i} \tr(e^{-\beta H_f}) = \frac{Z_f}{Z_i},
\end{equation}
Since $Z = e^{-\beta F}$, this yields  Eq.~(\ref{jarzynski}):
\begin{equation}\label{G_Jar}
G(i\beta) = \langle e^{-\beta W} \rangle = e^{-\beta \Delta F}.
\end{equation}
Notice that no assumptions have been made as to the speed of the process, so we conclude that the Jarzynski equality holds for a process arbitrarily far from equilibrium.

%
%
%
%
\section{\label{sec:resonance}Magnetic resonance}

\subsection{Statement of the problem}

We will now apply the concepts of the previous section to a magnetic resonance experiment.
The typical setup consists of a sample of non-interacting spin 1/2 particles (electrons, nucleons, etc.) placed under a strong static magnetic field $B_0$ in the $z$ direction.
The Hamiltonian of this interaction can be described using the Pauli matrix $\sigma_z$ as\cite{Sakurai2010}  
\begin{equation}\label{H0}
H_0 = - \frac{B_0}{2} \sigma_z.
\end{equation}
For simplicity, we choose to measure the field in energy units. 
Moreover, since $\hbar = 1$, the quantity $B_0$ also represents the characteristic precession frequency of the spin.
The eigenvalues of $H_0$ are $-B_0/2$ and $B_0/2$, corresponding to spin up and spin down respectively.

When the spin is coupled to a heat bath at a temperature $T$, its state will be given by the thermal density matrix $\rho_\text{th}$ in Eq.~(\ref{rho_thermal}). 
The Hamiltonian $H_0$ is already diagonal in the usual $|\pm\rangle$ basis which diagonalizes $\sigma_z$. 
Whenever this is true, the corresponding matrix exponential $e^{-H_0/T}$ may be easily computed by exponentiating the eigenvalues
\[
e^{-H_0/T} = \begin{pmatrix} e^{B_0/2T} & 0 \\ 0 & e^{- B_0/2T}\end{pmatrix}
\]
The partition function is the trace of this matrix, $Z = \tr(e^{-H_0/T})$,  and reads $Z = 2 \cosh(B_0/2T)$. 
The thermal density matrix $\rho_\text{th}  = e^{-H_0/T}/Z$ may then be written in a convenient way as 
\begin{equation}\label{rho_thermal2}
\rho_\text{th} =  \begin{pmatrix} \frac{1+f}{2} & 0 \\[0.2cm] 0 & \frac{1-f}{2} \end{pmatrix}.
\end{equation}
where $f$ is the equilibrium magnetization of the system 
\begin{equation}\label{sigma_z_thermal}
f = \langle \sigma_z \rangle_\text{th} = \tanh\left(\frac{B_0}{T}\right),
\end{equation}
corresponding to the paramagnetic response of a spin 1/2 particle. 

The work protocol is  implemented by applying a very small field of amplitude $B_1$ rotating in the $xy$ plane with frequency $\omega$. 
That is, the work parameter $\lambda$ is described here by the field $\bm{B}_1 = (B_1 \sin\omega t, B_1 \cos\omega t, 0)$. 
Typically,  $B_0 \sim 10$ T and $B_1 \sim 0.01$ T so we may always take as a good approximation that $B_1 \ll B_0$. 
The total Hamiltonian of the system now becomes 
\begin{equation}\label{H}
H(t) = -  \frac{B_0}{2} \sigma_z - \frac{B_1}{2}( \sigma_x \sin\omega t + \sigma_y \cos\omega t),
\end{equation}
The oscillating field plays the role of a perturbation which, albeit extremely weak, may nonetheless promote transitions between the up and down spin states. The transitions will be the most frequent at the resonance condition, which corresponds to $\omega = B_0$; ie, when the driving frequency $\omega$ is the same as the natural oscillation frequency $B_0$. 

To make progress, we must now compute the time evolution operator $U(t)$ defined in Eq.~(\ref{schrodinger}). 
Usually, accomplishing this for a time-dependent Hamiltonian is a very complicated task. 
Luckily, for the particular choice of $H$ in Eq.~(\ref{H}), this turns out to be quite simple. 
The first step is to define a new operator $\tilde{U}(t)$ from the relation
\begin{equation}\label{2U}
U(t) = e^{i \omega t \sigma_z/2} \tilde{U}(t).
\end{equation}
Substituting this in Eq.~(\ref{schrodinger}) one finds that $\tilde{U}$ must obey a modified Schr\"odinger equation
\begin{equation}\label{interaction_pic}
i \partial_t \tilde{U} = \tilde{H} \tilde{U},
\end{equation}
where 
\begin{equation}\label{Htilde}
\tilde{H} =  - \frac{(B_0 - \omega)}{2}  \sigma_z - \frac{B_1}{2}\sigma_y
\end{equation}
We therefore see that $\tilde{U}(t)$ evolves according to a  \emph{time-independent} Hamiltonian. 
Consequently,  the solution of Eq.~(\ref{interaction_pic}) will be simply $\tilde{U}(t)  = e^{-i \tilde{H} t}$ and the full time evolution operator will be given by 
\begin{equation}\label{U1}
U(t) = e^{i \omega t\sigma_z/2} e^{-i \tilde{H} t}
\end{equation}
It is important to notice that, since $\sigma_y$ and $\sigma_z$ do not commute, we cannot write $U(t)$ as $e^{i (\omega \sigma_z/2 - \tilde{H})t}$. 

It is also useful to have in hand an explicit formula for   $e^{-i \tilde{H} t}$. 
This can be accomplished using the following  trick.
Let $\mathcal{M}$ be an arbitrary matrix, which is such that $\mathcal{M}^2 = \mathbb{I}$ (where $\mathbb{I}$ is the $2\times2$ identity matrix). 
Then, if $\alpha$ is an arbitrary constant, a direct power series expansion of $e^{-i \alpha \mathcal{M}}$ yields
\begin{equation}\label{M_idea}
e^{-i\alpha \mathcal{M}} = \mathbb{I} \cos\alpha - i \mathcal{M} \sin\alpha.
\end{equation}
We can apply this idea to our problem by writing Eq.~(\ref{Htilde}) as 
\begin{equation}
\tilde{H} = -\frac{\Omega}{2} (\sigma_z \cos\theta + \sigma_y \sin\theta),
\end{equation}
where 
\begin{equation}\label{Omega}
\Omega = \sqrt{(B_0 -\omega)^2 + B_1^2 },\qquad \tan\theta = \frac{B_1}{B_0 - \omega}.
\end{equation}
Since $\sigma_i^2 = \mathbb{I}$, it follows that  $(\sigma_z \cos\theta + \sigma_y \sin\theta)^2 = \mathbb{I}$. 
Hence,  Eq.~(\ref{M_idea}) applies and we get
\begin{equation}
e^{-i \tilde{H} t} = \mathbb{I}\cos\left(\frac{\Omega t}{2}\right) + i (\sigma_z \cos\theta + \sigma_y \sin\theta) \sin\left(\frac{\Omega t}{2}\right)
\end{equation}

Inserting this result in Eq.~(\ref{U1}) we can finally write a closed (exact) formula for the full time evolution operator $U(t)$. 
After organizing the terms a bit, we get
\begin{equation}\label{U2}
U(t) = \begin{pmatrix} u(t) & v(t) \\[0.2cm] -v^*(t) & u^*(t) \end{pmatrix}
\end{equation}
where
\begin{eqnarray}
\label{u}u(t)&=&e^{i\omega t/2}\left[\cos\left(\frac{\Omega t}{2}\right) + i \cos\theta \sin\left(\frac{\Omega t}{2}\right)\right],\\[0.2cm]
\label{v}v(t)&=&e^{i\omega t/2} \sin\theta \sin\left(\frac{\Omega t}{2}\right),
\end{eqnarray}
We see that, apart from a phase factor $e^{i \omega t/2}$, 
the final result  depends only on $\Omega$ and $\theta$, which in turn  depend on $B_0$, $B_1$ and $\omega$ [cf. Eq.~(\ref{Omega})]. 

To understand the physics behind   $u(t)$ and $v(t)$, suppose the system starts in the pure state $ |+\rangle$. 
The probability that after a time $t$ it will be found in state $|-\rangle$ is $|\langle - | U(t) | + \rangle|^2$.
But looking at Eq.~(\ref{U2}) we see that $\langle - | U(t) | + \rangle = v(t)$.
Therefore, $|v|^2$ represents the  transition probability per unit time  for a jump to occur.
Moreover,  the unitarity condition $U^\dagger U = 1$  implies that $|v|^2 + |u|^2 = 1$, so $|u|^2$ is the probability that no transition occurs. 

From Eq.~(\ref{v}) we also see that $v \propto \sin \theta$. 
This therefore attributes a physical meaning to the angle $\theta$ [defined in Eq.~(\ref{Omega})] as representing the transition probability. 
It reaches a maximum precisely at resonance ($\omega = B_0$),  as we intuitively expect. 
In fact, at resonance  Eqs.~(\ref{u}) and (\ref{v}) simplify to 
\begin{eqnarray}
\label{ur}u(t)&=&e^{i\omega t/2} \cos\left(\frac{B_1 t}{2}\right),\\[0.2cm]
\label{vr}v(t)&=&e^{i\omega t/2}\sin\left(\frac{B_1 t}{2}\right)
\end{eqnarray}

Now that we have the initial density matrix [Eq.~(\ref{rho_thermal2})] and the time evolution operator [Eq.~(\ref{U2})], we may compute the evolution of any observable $A$ we wish using Eq.~(\ref{evolution_observable}). 
For instance, we could compute the evolution of the  magnetization components $\langle \sigma_i \rangle_t$. 
All calculations are reduced to the multiplication of $2\times 2$ matrices. 
For instance, the magnetization in the $z$ direction will be
\begin{equation}\label{sz}
\langle \sigma_z \rangle_t = f(1 - 2|v|^2) = f (\cos^2\theta + \sin^2\theta\cos\Omega t)
\end{equation}
where $f$ is given by Eq.~(\ref{sigma_z_thermal}) and we used the fact that $|u|^2 + |v|^2 = 1$.

\subsection{Average work}

When computing the expectation values of quantities related to the energy of the system, we may always use the unperturbed  Hamiltonian $H_0$ in Eq.~(\ref{H0}), instead of the full Hamiltonian $H(t)$ in Eq.~(\ref{H}), which is justified since $B_1 \ll B_0$. 
The energy of the system at any given time may therefore be found from Eq.~(\ref{evolution_observable}) with $A = H_0$. 
It reads
\[
\langle H_0 \rangle_t  = - \frac{B_0}{2} \langle \sigma_z \rangle_t  = - \frac{B_0 f}{2}(1-|v|^2)
\] 
The average work at time $t$ is then simple the  difference between the energy at time $t$ and the energy at time $0$:
\begin{equation}\label{ave_work}
\langle W \rangle_t = f B_0 |v|^2 =  f B_0\; \frac{ B_1^2}{\Omega^2} \; \sin^2\left(\frac{\Omega t}{2}\right),
\end{equation}
where we recall that $\Omega = \sqrt{(B_0 -\omega)^2 + B_1^2}$.
The average work therefore oscillates indefinitely with  frequency $\Omega/2$.
This is a consequence of the fact that the time evolution is unitary. 

The amplitude multiplying the average work is proportional to the initial magnetization $f$ and to the ratio  $B_1^2/[(B_0 -\omega)^2 + B_1^2]$. 
This is known as a Lorentzian function. It presents a sharp peak at the resonance frequency $\omega = B_0$, which becomes sharper the smaller is the value of $B_1$.
The maximum possible work therefore occurs at resonance and has the value $f B_0$.

The equilibrium free energy, Eq.~(\ref{F}) is  $F = -T \ln Z$, where $Z = 2 \cosh(B_0/2T)$. 
Thus, the  free energy of the initial state (at time $t = 0$) and the final state (at any arbitrary time $t$) are the same: $\Delta F = 0$. 
This is a consequence of the fact that $B_1 \ll B_0$. 
According to Eq.~(\ref{second_law2}) we should then expect $\langle W \rangle \geq 0$, which is indeed observed in Eq.~(\ref{ave_work}).

\subsection{Characteristic function and distribution of work}


Next we turn to the characteristic function $G(r)$ given in Eq.~(\ref{G2}), with both $H_i$ and $H_f$ replaced by $H_0$ in Eq.~(\ref{H0}). 
After carrying out the matrix multiplications, we get the following very simple formula:
\begin{equation}\label{G3}
G(r) = |u(t)|^2 + |v(t)|^2 \bigg\{\frac{(1+f)}{2} e^{i B_0 r} + \frac{(1-f)}{2} e^{-i B_0 r} \bigg\}
\end{equation}
If we set $r = i\beta$ and recall the definition of $f$ in Eq.~(\ref{sigma_z_thermal}), we find that the term inside brackets becomes 1. 
Thus,  we are  left with $\langle e^{-\beta W} \rangle = G(i\beta) = |u|^2 + |v|^2 = 1$. 
This is the Jarzynski equality, Eq.~(\ref{G_Jar}), since $\Delta F = 0$.

Expanding $G(r)$ in a power series, as in Eq.~(\ref{G_series}), we also obtain the statistical moments of the work. 
The first order term will give $\langle W \rangle$ exactly as in Eq.~(\ref{ave_work}). 
Similarly, the second moment can easily be found to be $\langle W^2 \rangle = B_0^2 |v|^2$.
As a consequence, the variance of the work is 
\begin{equation}\label{varW}
\text{var}(W) = \langle W^2 \rangle - \langle W \rangle^2 = B_0^2 |v|^2 (1 - f^2 |v|^2)
\end{equation}

Finally, we may compute the full distribution of work $P(W)$. 
The simplest way to do so is through the  characteristic function. 
Recall from Eq.~(\ref{PG}) that $P(W)$ is the inverse Fourier transform of $G(r)$. 
To carry out the computation, we must use the following representation for the Dirac delta function:
\begin{equation}\label{delta}
\frac{1}{2\pi}\int\limits_{-\infty}^\infty e^{i (a-b) r} \ud r = \delta(a-b).
\end{equation}
Using this in Eq.~(\ref{G3}) we then find
\begin{IEEEeqnarray}{rCl}
\label{PWres} P(W) = |u|^2 \delta (W) &&+ |v|^2 \frac{1+f}{2} \delta(W-B_0) \\[0.2cm]
&&+ |v|^2 \frac{1-f}{2} \delta(W+B_0), \nonumber
\end{IEEEeqnarray}
We therefore see that the work, interpreted as a random variable, may take on three distinct values: $W = 0, +B_0$ or $-B_0$. 

The physics behind this result is the following. 
Looking back at the original Hamiltonian in Eq.~(\ref{H0}), we see that $B_0$ is the energy spacing between the up and down states.
The event where $W = + B_0$ corresponds to the situation where the spin was originally up and then flipped down  (``up-down flip'').
The change in energy in this case is  $B_0/2 - (-B_0/2) = B_0$. 
Similarly, $W = -B_0$ corresponds to a down-up flip.
And, finally, $W = 0$ corresponds to no flip at all.

We may also find the distribution $P(W)$ ``by hand'', using  Eq.~(\ref{PW}). 
For instance, the value $W = +B_0$ corresponds to the up-down flip.
The initial probability to have a particle up is $(1+f)/2$ and the transition rate is $|\langle - | U(t) | + \rangle|^2  = |v|^2$. 
Thus, $P(W = B_0) = |v|^2 (1+f)/2$, which agrees with Eq.~(\ref{PWres}).
The other two probabilities may be computed in an identical way. 

From the second law we expect that $W > 0$.
But our results show that in a down-up flip we should have  $W  =- B_0$. 
Hence,  $P(W = -B_0)$ is the probability of observing a  local violation of the second law. 
On the other, hand, notice in Eq.~(\ref{PWres}) that $P(W = \pm B_0)$ is proportional to $1\pm f$ where $f = \tanh(\beta B_0)$.
Thus, up-down flips are always more likely than down-up flips, ensuring that $\langle W \rangle \geq 0$.
That is to say, violations to the second law are always the exception, never the rule. 

It is also possible to express Eq.~(\ref{PWres})  in terms only of the magnetization $\langle \sigma_z \rangle_t$,  Eq.~(\ref{sz}). 
This is interesting because $\langle \sigma_z \rangle_t$ is a quantity that can be directly accessed experimentally. 
Substituting $|v|^2 = (f-\langle \sigma_z\rangle_t)/2f$ in Eq.~(\ref{PWres}) gives
\begin{equation}\label{Prob_exp}
\text{Prob}(W = \pm B_0) = \left(\frac{f - \langle \sigma_z \rangle_t}{2f}\right)\frac{1\pm f}{2}
\end{equation}
This formula shows that by measuring the average magnetization of a system, which is a macroscopic observable, we may extract the full distribution of work for a single spin 1/2 particle. 
The Jarzynski equality for a quantum system was first confirmed experimentally in Ref.~\onlinecite{Batalhao2014} also using magnetic resonance. 
However, in their experiment it was necessary to use two interacting spins, which turns out to be a consequence of the fact that in their case $[H_i, H_f]\neq 0$. 
In our case, since $B_1 \ll B_0$, the initial and final Hamiltonians commute, thus we are able to relate the distribution of work to the properties of a single spin.

%
%
%
%
\subsection{\label{sec:many}Statistics of the work performed on a large number of particles}

So far we have studied the work performed by an external magnetic field on  a single spin 1/2 particle.
It is a remarkable fact that with recent advances in experimental techniques  it is now possible to experiment with just a single particle.
Notwithstanding, in most situations one is still usually faced with a system containing a large number of particles. 
The next natural step is therefore to consider the work performed on $N$ spin 1/2 particles. 
For simplicity, we will assume that the particles do not interact (otherwise the problem would be much more difficult to deal with).

The work corresponds to energy differences and for non-interacting systems, energy is an additive quantity. 
Hence, the total work $\mathcal{W}$ performed during a certain process will be the sum of the work performed on each individual particle:
\begin{equation}\label{WN}
\mathcal{W} = W_1 + \ldots+W_N.
\end{equation}
Since all spins are independent, it follows from this result that $\langle \mathcal{W} \rangle = N \langle W \rangle$, where $\langle W \rangle$ is the average work  in Eq.~(\ref{ave_work}).
This is  a manifestation of the fact that work, as with energy, is an extensive quantity. 

The problem has thus been reduced to the sum of independent and identically distributed random variables, something which is discussed extensively in introductory probability courses (Ref.~\onlinecite{Ross2010}, chapter 6).
It is in problems such as this that the characteristic function shows its true power. 
Since the variables are statistically independent, we have from Eq.~(\ref{WN}) that
\begin{equation}
\langle e^{i r \mathcal{W}} \rangle  = \langle e^{i r (W_1+\ldots+W_N)} \rangle = \langle e^{i r W_1} \rangle \ldots \langle e^{i r W_N} \rangle.
\end{equation}
Moreover, since all spins are identical, each term on the right-hand side corresponds exactly to the characteristic function $G(r)$ in Eq.~(\ref{G3}). 
Thus, the characteristic function of the total work $\mathcal{W}$ will be 
\begin{equation}\label{mathG}
\mathcal{G}(r) = \langle e^{i r \mathcal{W}} \rangle = G(r)^N.
\end{equation}
%

Referring  back to Eq.~(\ref{G3}), let us introduce momentarily the notation $b_0 = |u|^2$ and $b_\pm = |v|^2 (1\pm f)/2$.
Then Eq.~(\ref{mathG}) may be written as 
\begin{equation}\label{mathG2}
\mathcal{G}(r) = \bigg(b_0 + b_+ e^{i B_0 r} + b_- e^{-i B_0 r}\bigg)^N.
\end{equation}
The importance of this formula lies in its connection with the probability distribution $P(\mathcal{W})$, established via the inverse Fourier transform [Eq.~(\ref{PG})]. 
If we expand the product in Eq.~(\ref{mathG2}) we will get 
\begin{equation}\label{mathG3}
\mathcal{G}(r) = \sum\limits_{k = -N}^N \Gamma_k e^{i r B_0 k},
\end{equation}
where $\Gamma_k$ are complicated combinations of  the $b$ coefficients, which result from expanding Eq.~(\ref{mathG2}). 
When we take the inverse Fourier transform this  is mapped into 
\begin{equation}\label{PmathW}
P(\mathcal{W}) = \sum\limits_{k = -N}^N \Gamma_k \; \delta(\mathcal{W} - B_0 k).
\end{equation}
We therefore see that $\mathcal{W}$ may take on values between $-N B_0$ and $N B_0$. 
A work of $N B_0$ for instance, corresponds to an event where all spins have flipped from up to down. 
Similarly, a work of $(N-1) B_0$ corresponds to $N-1$ spins flipping. 
And there are $N$ possibilities for which of the spins did not flip. 
For other values the situation becomes even more complicated.

The  characteristic function~(\ref{mathG2}) may also be interpreted as a random walk with $N$ steps. 
In a single step, one may think of $b_+$ and $b_-$ as the probability of taking a step to the right or to the left (while $b_0$ is the probability of not moving).
If we repeat this $N$ times, we get a characteristic function of the form~(\ref{mathG2}). 
Moreover, $P(\mathcal{W})$ plays the role of the distribution of discrete positions of the random walk.

\begin{figure}[!t]
\centering
\includegraphics[width=0.22\textwidth]{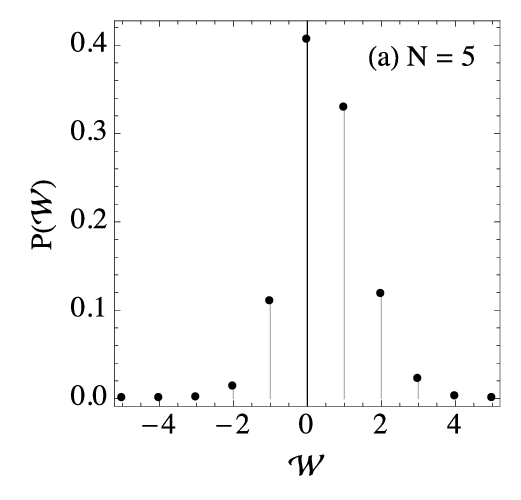}\quad
\includegraphics[width=0.22\textwidth]{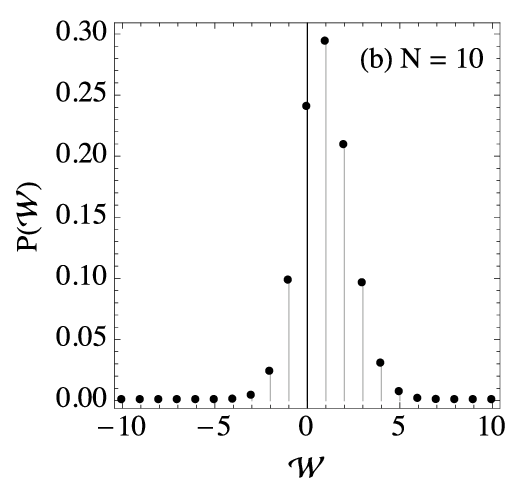}\\
\includegraphics[width=0.22\textwidth]{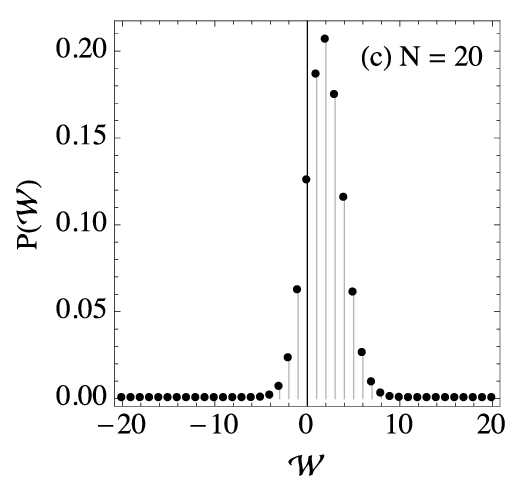}\quad
\includegraphics[width=0.22\textwidth]{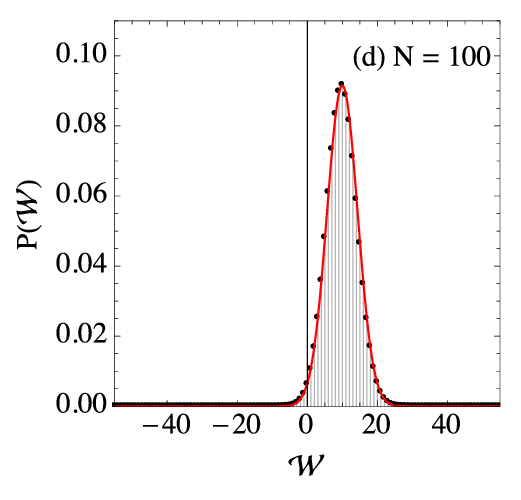}
\caption{\label{fig:PW}The distribution $P(\mathcal{W})$ computed from Eq.~(\ref{PmathW}) for different system sizes $N$. 
The parameters used in this plot were $B_0 = 1$, $\omega = 0.8$, $B_1 = 0.1$, $f = 0.5$ and $t = \pi/\Omega$.
The solid line in image (d) is a Gaussian distribution with the same mean and variance.
}
\end{figure}

Eq.~(\ref{PmathW}) is illustrated in Fig.~\ref{fig:PW} for an arbitrary choice of parameters, as explained in the figure caption. 
The important point to be drawn from this analysis is that there is a certain finite probability to observe  a \emph{negative} work $\mathcal{W}$. 
Since $\Delta F = 0$, these would then correspond to local violations of the second law. 
However, notice also that as the size $N$ increases, the relative probability that $\mathcal{W}<0$ diminishes quickly. 
In fact, a more detailed analysis shows that
\begin{equation}
\text{Prob}(\mathcal{W} < \Delta F) \sim e^{- N}
\end{equation}
Thus, if the sample is  macroscopic, it becomes extremely unlikely to observe such a violation. 
This is why, in our everyday experience, the second law is always satisfied. 

Lastly, we should mention that when $N$ is large we may approximate $P(W)$ by a Gaussian distribution, as a consequence of the central limit theorem.\cite{Ross2010}
This distribution will have mean $\langle \mathcal{W} \rangle = N \langle W \rangle$ and 
variance $\text{var}(\mathcal{W}) = N \text{var}(W)$, which are quantities we already know [Eqs.~(\ref{ave_work}) and (\ref{varW})]. 
This distribution is plotted as a solid line in Fig.~\ref{fig:PW}(d), to be compared with the exact solution.

%
%
%
%
\section{\label{sec:conc}Conclusions}

The goal of this paper was to introduce the student to the concepts of quantum thermodynamics and work fluctuations.
This area of research is very active and lies at the boundary between many well established areas, such as non-equilibrium statistical mechanics, quantum physics and condensed matter. 
It also has deep connections with quantum information and quantum computing. 
Notwithstanding, unlike most frontier areas, its basic ideas  can be understood using only   concepts learned in undergraduate quantum and statistical mechanics courses. 
This makes for a unique opportunity to put the student up to speed with the current research, which was the goal of this paper. 

In this paper we have aimed to give an introduction which was as simple as possible but, at the same time,  useful for students entering this area and for a general audience. 
We would like to take as an example  the characteristic function. 
This is a concept which is not necessary,  \emph{per se}, to understand the ideas involved in this area. 
But it is such a useful  concept  that, we feel, every student in this area should know how to work with it. 
Hence, even though the use of the characteristic function may have complicated the analysis a bit, we strongly believe that it was worth it.

In this paper we have made the assumption that the motion of the system is unitary. 
That is, that during the time evolution we may assume that the system is not connected to a heat reservoir. 
This is certainly true for many systems, including nuclear magnetic resonance experiments. 
However, in many other scenarios it becomes important to consider also \emph{open quantum systems}. 
That is, systems whose dynamics evolve coupled to a heat bath. 
The tools of open quantum systems are already extensively used in many areas, but its role in quantum thermodynamics is still in its infancy. 
We believe that in the future it will play a particularly important role in the developments of this area.

\begin{acknowledgements}
The authors would like to thank Prof. M\'ario Jos\'e de Oliveira, Prof. Ivan Santos Oliveira and Prof. Andr\'e Timpanaro for fruitful discussions.
For their financial support, the authors would like to acknowledge the Brazilian funding agencies FAPESP  (2014/01218-2) and CNPq (307774/2014-7).
Support from the INCT of Quantum Information is also acknowledged.
\end{acknowledgements}

%

\end{document}